\documentclass[journal,comsoc]{IEEEtran}
\usepackage[T1]{fontenc}% optional T1 font encoding

\ifCLASSINFOpdf
  \usepackage[pdftex]{graphicx}
  \graphicspath{{../pdf/}{../jpeg/}}
  \DeclareGraphicsExtensions{.pdf,.jpeg,.png}
\else
  \usepackage[dvips]{graphicx}
  \graphicspath{{../eps/}}
  \DeclareGraphicsExtensions{.eps}
\fi

\usepackage{amsmath}
\usepackage{stfloats}
\usepackage{bm}
\usepackage{algorithm}  
\usepackage{algorithmicx}  
\usepackage{algpseudocode}

\usepackage{url}

\begin{document}
\bstctlcite{BSTcontrol}
% paper title

\title{Energy Efficiency Maximization for STAR-RIS Assisted Full-Duplex Communications}
\author{
	\IEEEauthorblockN{Pengxin Guan, Yiru Wang, Hongkang Yu and Yuping Zhao}
	%\thanks{This work was supported by Pengcheng National Laboratory No.PCL2021A04 of China. \emph{(Corresponding author: Yuping Zhao.)}}
	\thanks{Pengxin Guan, Yiru Wang, Hongkang Yu, and Yuping Zhao are with the School of Electronics Engineering and Computer Science, Peking University, Beijing, China. (email: yuping.zhao@pku.edu.cn).}
	
	%\IEEEauthorblockA{\{guanpengxin, yuhongkang, yuping.zhao\}@pku.edu.cn}

	%{Corresponding author: Yuping Zhao (e-mail: yuping.zhao@pku.edu.cn).}
}

% The paper headers
\markboth{Journal of \LaTeX\ Class Files,~Vol.~14, No.~8, August~2015}%
{Shell \MakeLowercase{\textit{et al.}}: Bare Demo of IEEEtran.cls for IEEE Communications Society Journals}

% make the title area
\maketitle

% As a general rule, do not put math, special symbols or citations
% in the abstract or keywords.
\begin{abstract}
This letter investigates a novel simultaneous transmission and reflection reconfigurable intelligent surface (STAR-RIS) aided full-duplex (FD) communication system. A FD base station (BS) communicates with an uplink (UL) user and a downlink (DL) user simultaneously over the same time-frequency dimension assisted by a STAR-RIS. We aim to maximize the system energy efficiency by jointly optimizing the transmit power of the BS and UL user, and passive beamforming at the STAR-RIS. We decouple the non-convex problem into two subproblems and optimize them iteratively under the alternating optimization framework. The Dinkelbach’s method is used to solve the power optimization subproblem, while the penalty-based method and successive convex approximation (SCA) are applied to design passive beamforming at the STAR-RIS. Simulation results demonstrate the superior performance of our proposed scheme compared to other baseline schemes.
\end{abstract}

% Note that keywords are not normally used for peerreview papers.
\begin{IEEEkeywords}
STAR-RIS, full-duplex, energy efficiency, alternating optimization.
\end{IEEEkeywords}

% For peer review papers, you can put extra information on the cover
% page as needed:
% \ifCLASSOPTIONpeerreview
% \begin{center} \bfseries EDICS Category: 3-BBND \end{center}
% \fi
%
% For peerreview papers, this IEEEtran command inserts a page break and
% creates the second title. It will be ignored for other modes.
\IEEEpeerreviewmaketitle

\section{Introduction}

\IEEEPARstart{R}{ecently}, reconfigurable intelligent surface (RIS) has become a promising technology for the sixth generation (6G) wireless communications, which attracts increasing attention from the academia and industry \cite{ref1}. However, since conventional RIS can only reflect incident signals, the transmitter and receiver need to be deployed on the same side of the RIS, which limits the flexibility of RISs. Fortunately, a novel simultaneous transmission and reflection reconfigurable intelligent surface (STAR-RIS) was proposed to improve the convenience of communication \cite{ref2}, \cite{ref3}. The STAR-RIS can achieve $360^{\rm{o}}$ coverage by spliting the incident signal into transmission (T) region and reflection (R) region \cite{ref4}. 

By properly designing the transmission and reflection coefficients at the STAR-RIS, the system performance can be further improved. In \cite{ref2}, the author studied the power minimization problem for STAR-RIS aided downlink communication system. The STAR-RIS assisted secure nonorthogonal multiple access communications was also be investigated in \cite{ref5}. In \cite{ref6}, the author maximized the weighted sum secrecy rate in a STAR-RIS aided secrecy networks. The athour maximized the weighted sum rate in STAR-RIS aided multiple-input multiple-output system in \cite{ref7}.

\begin{figure}%[!t]
	\centering
	\includegraphics[width=2.92in]{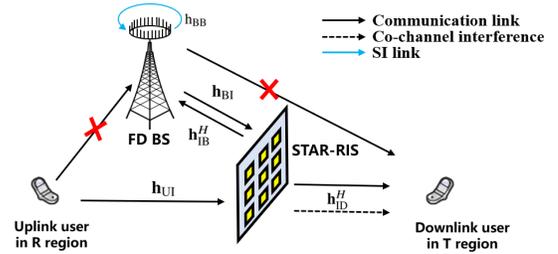}
	% where an .eps filename suffix will be assumed under latex,
	% and a .pdf suffix will be assumed for pdflatex; or what has been declared
	% via \DeclareGraphicsExtensions.
	\caption{A STAR-RIS assisted FD communication system.}
\end{figure}
However, the literature mentioned above only studied the half-duplex (HD) systems, which cause the spectral efficiency (SE) loss. The full-duplex (FD) technology enables signal transmission and reception over the same time-frequency dimension and thus can double the SE theoretically compared with HD \cite{ref8}. Nevertheless, communications in FD mode would suffer from strong self-interference (SI) signal. Fortunately, a number of SI cancellation (SIC) methods can suppress the SI power to the noise floor, which promotes the FD-based applications \cite{ref9}. Hence, the combination of FD and STAR-RIS will further benefit wireless communication systems.

In this letter, we consider a STAR-RIS aided FD system, where a FD base station (BS) communicates with an uplink (UL) user and a downlink (DL) user simultaneously assisted by a STAR-RIS. We aim to maximize the system energy efficiency (EE), while satisfying both the UL and DL minimum rate constraints. To the best of our knownledeg, similar work has not yet been discussed in the literature. The main contributions are summarized as follows:
\begin{itemize}
	\item We are the first to consider the energy efficiency maximization for STAR-RIS aided FD system by jointly optimizing the transmit power of the BS and UL user, as well as the passive beamforming at the STAR-RIS. 
	\item We decouple the non-convex problem into transmit power optimization and passive beamforming design subproblems and use the alternative optimization (AO) to solve them iteratively. The Dinkelbach’s method is used to solve the power optimization subproblem while the penalty-based method and successive convex approximation (SCA) are applied to design passive beamforming at the STAR-RIS.
	\item Simulation results show the benifits of combining the STAR-RIS and FD communication in terms of EE compared with other benchmark schemes.
\end{itemize}

\emph{Notation}: $\left| x \right|$ denote the absolution value of a scalar $x$. Tr(${\bf{X}}$), ${{\bf{X}}^H}$, ${\rm{\uplambda }}({\bf{X}})$, and rank(${\bf{X}}$) denote the trace, conjugate transpose, maximum eigenvalue, and rank of the matrix ${\bf{X}}$, respectively. Diag(${\bf{x}}$) is a diagonal matrix with the entries of ${\bf{x}}$ on its main diagonal. diag(${\bf{X}}$) is a vertor whose elements are extracted form the main diagonal elements of matrix ${\bf{X}}$. ${\mathbb{C}^{m \times n}}$ denotes the space of ${m \times n}$ complex matrices. 
\section{SYSTEM MODEL AND PROBLEM FORMULATION}

\subsection{System Model Description}
As shown in Fig. 1, we consider a FD BS serves an UL user in R region and a DL user in T region simultaneously assisted by a STAR-RIS over the same time-frequency dimension. The BS is equipped with a single transmit antenna and a single receive antenna, and both UL and DL are equipped with a single antenna. We assume that there is no direct link between the users and BS because of deep fading or heavy shadowing. 

The STAR-RIS, denoted by I, has $M$ passive elements. We assume the STAR-RIS adopts the energy splitting (ES) protocol \cite{ref2}, where all elements can simultaneously transmit and reflect signals. The transmission and reflection coefficients vectors of the STAR-RIS can be expressed as ${{\bf{q}}_t} = {(\sqrt {\beta _1^t} {e^{j\phi _1^t}},...,\sqrt {\beta _M^t} {e^{j\phi _M^t}})^H}$ and ${{\bf{q}}_r} = {(\sqrt {\beta _1^r} {e^{j\phi _1^r}},...,\sqrt {\beta _M^r} {e^{j\phi _M^r}})^H}$, where $\sqrt {\beta _m^t}  \in \left[ {0,1} \right]$, $\phi _m^t \in \left[ {0,2\pi } \right)$ and $\sqrt {\beta _m^r}  \in \left[ {0,1} \right]$, $\phi _m^r \in \left[ {0,2\pi } \right)$ denote the amplitude and phase shift of the $m$-th element’s T\&R coefficients. To obey the law of energy conservation \cite{ref2}, we restrict that $\beta _m^t + \beta _m^r = 1$. We assume an ideal STAR-RIS with adjustable surface electric and magnetic impedance is deployed in the system, thus $\phi _m^t$ and $\phi _m^r$ can be tuned independently \cite{ref2}, \cite{ref3}.

Denote ${{\bf{h}}_\text{UI}} \in {{\mathbb C}^{M \times 1}}$, ${\bf{h}}_\text{IB}^H \in {{\mathbb C}^{1 \times M}}$, ${{\bf{h}}_\text{BI}} \in {{\mathbb C}^{M \times 1}}$, ${\bf{h}}_\text{ID}^H \in {{\mathbb C}^{1 \times M}}$ as the channel between the UL user and the STAR-RIS, between the STAR-RIS and the BS, between the BS and the STAR-RIS, between the STAR-RIS and the DL user respectively. ${{\text{h}}_\text{BB}} \sim \rm{\mathcal{CN}} (0,\sigma _{{\rm{SI}}}^2)$ represents the residual SI (RSI) channel coefficient at BS \cite{ref10}.

Then, the signal received at the BS can be expressed as
\begin{equation}
{y_U} = {\bf{h}}_\text{IB}^H{{\bm{\Phi }}_r}{{\bf{h}}_\text{UI}}\sqrt {{p_U}} {x_U} +  {{\text{h}_\text{BB}}\sqrt {{p_D}} {x_D}} + {n_U},
\end{equation}
where $p_{U}$ and $p_{D}$ are transmit power of UL user and BS, $x_{U}$ and $x_{D}$ are the independent data symbol with normalized power of UL and DL, ${{\bf{\Phi }}_r} = \text{Diag}({{\bf{q}}_r})$, and ${n_U} \sim \rm{\mathcal{CN}} (0,\sigma _U^2)$ is the additive white Gaussian noise (AWGN) at the BS. 

The received signal at the DL user can be expressed as
\begin{equation}\label{yd}
{y_D} = {\bf{h}}_\text{ID}^H{{\bm{\Phi }}_t}{{\bf{h}}_\text{BI}}\sqrt {{p_D}} {x_D} +  {{\bf{h}}_\text{ID}^H{{\bm{\Phi }}_t}{{\bf{h}}_\text{UI}}\sqrt {{p_U}} {x_U}} + {n_D},
\end{equation}
where ${{\bf{\Phi }}_t} = \text{Diag}({{\bf{q}}_t})$, and ${n_D} \sim \rm{\mathcal{CN}}(0,\sigma _D^2)$ is the AWGN at the DL user. Note that the second term of equation (2) is co-channel interference due to the FD mode.

Therefore, the achievable data rate in bits second per Hertz (bps/Hz) of the UL and DL can be formulated as 
\begin{equation}\label{ru}
{R_U} = {\log _2}\left( {1 + \frac{{{p_U}{{\left| {{\bf{h}}_\text{IB}^H{{\bm{\Phi }}_r}{{\bf{h}}_\text{UI}}} \right|}^2}}}{{{p_D}{{\left| {{\text{h}_\text{BB}}} \right|}^2} + \sigma _U^2}}} \right),
\end{equation}
and
\begin{equation}\label{rd}
{R_D} = {\log _2}\left( {1 + \frac{{{p_D}{{\left| {{\bf{h}}_\text{ID}^H{{\bm{\Phi }}_t}{{\bf{h}}_\text{BI}}} \right|}^2}}}{{{p_U}{{\left| {{\bf{h}}_\text{ID}^H{{\bm{\Phi }}_t}{{\bf{h}}_\text{UI}}} \right|}^2} + \sigma _D^2}}} \right).
\end{equation}

The total power consumption of system can be given as
\begin{equation}
{P_{tot}} = {P_c} + \frac{1}{\rho }\left( {{p_U} + {p_D}} \right) + M{P_s}+{P_{\rm{SIC}}},
\end{equation}
where ${P_s}$ is the power consumption of each element at STAR-RIS and $\rho $ is the power amplifier efficiency. ${P_{{\rm{SIC}}}} = \xi {p_D} + {P_{c0}}$ represents the SIC power consumption, where $\xi$ is the isolation factor characterizing the effect of propagation domain SIC and ${P_{c0}}$ is the power consumption induced by the SIC circuits \cite{ref11}. ${P_c}$ is other circuit power consumption in the system.
%We define the ratio between the achievable sum rate and the total power consumption as the energy efficiency (EE), which can be expressed as
%\begin{equation}
%\eta  = \frac{R}{{{P_{tot}}}},
%\end{equation}
%where $R = {R_U} + {R_D}$ is the achievable sum rate of the system.
\subsection{Problem Formulation}
Our objective is to maxmize the EE by jointly designing the transmit power of BS and UL user, and passive beamforming at the STAR-RIS, subject to the minimal data requirement of the UL and DL, and transmit power budget constraints. Mathematically, the optimization problem is formulated as
\begin{subequations}\label{opt1}
	\begin{alignat}{2} 
		\mathcal{P}1:\quad& {\mathop {\max}\limits_{{{p_{U}}},{{p_{D}}}, {{\bf{q}}_l}}} &\ &\eta=\frac{R}{{{P_{tot}}}} \label{opt1A} \\
		& \quad{\textrm {s.t.}}
		&&{R_U} \geqslant R_U^{\text{th}},\label{opt1B}\\
		&&&{R_D} \geqslant R_D^{\text{th}},\label{opt1C}\\
		&&&0 \le {p_k} \le p_k^{\max },{\rm{ }}\forall k \in \{ U,D\},\label{opt1D}\\
		&&&{{\bf{q}}_l} \in \mathcal{F},\forall l \in \left\{ {t,r} \right\}, \label{opt1E}
	\end{alignat}
\end{subequations}	
where $R = {R_U} + {R_D}$ is the achievable sum rate of the system, $R_U^{\rm{th}}$ and $R_D^{\rm{th}}$ denote the minimum data rate of the UL and the DL respectively. $p_k^{\max },{\rm{ }}\forall k \in \{ U,D\}$ is the corresponding maximum transmit power. $\mathcal{F}$ denotes the feasible set for the STAR-RIS T\&R coefficient, in which $\sqrt {\beta _m^t}  \in \left[ {0,1} \right]$, $\sqrt {\beta _m^r}  \in \left[ {0,1} \right]$, $\beta _m^t + \beta _m^r = 1$, $\phi _m^t \in \left[ {0,2\pi } \right)$ and $\phi _m^r \in \left[ {0,2\pi } \right)$.
\section{TRANSMIT POWER MINIMIZATION ALGORITHM DESIGN}
In this section, we decouple the problem $\mathcal{P}1$ into power optimization and passive beamforming design subproblems, and adopt the AO method to solve them iteratively.

\subsection{Power Optimization With Given ${{\bf{q}}_t}$ and ${{\bf{q}}_r}$}
\newcounter{TempEqCnt}                         % 创建临时变量TempEqCnt
\setcounter{TempEqCnt}{\value{equation}} % 将当前公式序号 赋给TempEqCnt
\setcounter{equation}{6}                           % 当前公式序号变为x，x等于长公式应有的序号减1.
\begin{figure*}[ht]
	\begin{equation}
	R\left( {{p_U},{p_D}} \right) = \underbrace {{{\log }_2}\left( {{p_D}{\gamma _{BB}} + \sigma _U^2 + {p_U}{\gamma _{1}}} \right)}_{{f_1}\left( {{p_D},{p_U}} \right)}{\text{ + }}\underbrace {{{\log }_2}\left( {{p_U}{\gamma _{3}} + \sigma _D^2 + {p_D}{\gamma _{2}}} \right)}_{{f_2}\left( {{p_D},{p_U}} \right)} - \underbrace {{{\log }_2}\left( {{p_D}{\gamma _{BB}} + \sigma _U^2} \right)}_{{f_3}\left( {{p_D}} \right)} - \underbrace {{{\log }_2}\left( {{p_U}{\gamma _{3}} + \sigma _D^2} \right)}_{{f_4}\left( {{p_U}} \right)} .
	\end{equation}\hrulefill
\end{figure*}
\setcounter{equation}{\value{TempEqCnt}} % 把TempEqCnt中存的公式序号赋回给当前公式序号
With given ${{\bf{q}}_t}$ and ${{\bf{q}}_r}$, $\mathcal{P}1$ can be transformed into power optimization subproblem as follows
\setcounter{equation}{7} 
\begin{subequations}\label{opt2}
	\begin{alignat}{2} 
	\mathcal{P}2:\quad& {\mathop {\max}\limits_{{{p_U},{p_D}}}} \quad\frac{{R\left( {{p_U},{p_D}} \right)}}{{{P_{tot}}\left( {{p_U},{p_D}} \right)}} \label{opt2A} \\
	{\textrm {s.t.}}
	&\quad 0 \le {p_k} \le p_k^{\max },{\rm{ }}\forall k \in \{ U,D\},\label{opt2B}\\
	&\quad{p_U}{\left| {{\bf{h}}_{{\rm{IB}}}^H{{\bf{\Phi }}_r}{{\bf{h}}_{{\rm{UI}}}}} \right|^2} \ge R_U^{{\rm{TH}}}({p_D}{\left| {{{\rm{h}}_{{\rm{BB}}}}} \right|^2} + \sigma _U^2),\label{opt2C}\\
	&\quad{p_D}{\left| {{\bf{h}}_{{\rm{ID}}}^H{{\bf{\Phi }}_t}{{\bf{h}}_{{\rm{BI}}}}} \right|^2} \ge R_D^{{\rm{TH}}}({p_U}{\left| {{\bf{h}}_{{\rm{ID}}}^H{{\bf{\Phi }}_t}{{\bf{h}}_{{\rm{UI}}}}} \right|^2} + \sigma _D^2),\label{opt2D}
	\end{alignat}
\end{subequations}
where ${R\left( {{P_U},{P_D}} \right)}$ is shown at the top of the next page, ${{\bf{h}}_1} = \text{Diag}({\bf{h}}_\text{IB}^H){{\bf{h}}_\text{UI}}$, ${{\bf{h}}_2} = \text{Diag}({\bf{h}}_\text{ID}^H){{\bf{h}}_\text{BI}}$, ${{\bf{h}}_3} = \text{Diag}({\bf{h}}_\text{ID}^H){{\bf{h}}_\text{UI}}$, ${\gamma _1} = {\left| {{\bf{q}}_r^H{{\bf{h}}_1}} \right|^2}$, ${\gamma _2} = {\left| {{\bf{q}}_t^H{{\bf{h}}_2}} \right|^2}$, ${\gamma _3} = {\left| {{\bf{q}}_t^H{{\bf{h}}_3}} \right|^2}$, ${\gamma _{BB}} = {\left| {{{\rm{h}}_{{\rm{BB}}}}} \right|^2}$, $R_U^\text{TH} = {2^{R_U^\text{th}}} - 1$ and $R_D^\text{TH} = {2^{R_D^\text{th}}} - 1$. 

By applying the Dinkelbach method \cite{ref12}, we can transform $\mathcal{P}2$ into the following optimization problem:
\begin{subequations}\label{opt3}
	\begin{alignat}{2} 
	\mathcal{P}2^{\prime}:\quad& {\mathop {\max}\limits_{{{p_U},{p_D}}}} &\ &R({p_U},{p_D}) - \alpha {P_{tot}}({p_U},{p_D}) \label{opt3A} \\
	& \quad{\textrm {s.t.}}
	&&\text{8(b)-8(d)},\label{opt3B}
	\end{alignat}
\end{subequations}
where $\alpha \ge 0$ is the Dinkelbach variable. According to \cite{ref12}, $\alpha $ can be iteratively updated, which can be given as
\begin{equation}
{\alpha ^{(n)}} = \frac{{R(p_U^{(n)},p_D^{(n)})}}{{{P_{tot}}(p_U^{(n)},p_D^{(n)})}}
\end{equation}
where $n$ is the iteration index. ${p_U^{(n)}}$ and ${p_D^{(n)}}$ represent the UL and DL transmit power at $n$-th iteration resprectively and ${\alpha ^{(n)}}$ is the corresponding EE of system.

However, problem $\mathcal{P}2^{\prime}$ is still non-trival with given $\alpha$ due to non-concave objective function (9a). Since ${{f_3}\left( {{P_D}} \right)}$ and ${{f_4}\left( {{P_U}} \right)}$ are differentiable concave functions, we can utilize the first-order Taylor approximation at local point in the $n$-th iteration to approximate it, which can be written as follow
\begin{equation}
\begin{aligned}
{f_3}\left( {{p_D}} \right) \leqslant {f_3}\left( {\left. {{p_D}} \right|p_D^{\left( n \right)}} \right) &= \frac{1}{{\text{ln2}}} \cdot \frac{{{\gamma _{BB}}}}{{p_D^{\left( n \right)}{\gamma _{BB}} + \sigma _U^2}}\left( {{p_D} - p_D^{\left( n \right)}} \right) \hfill \\
&\quad+ {\log _2}\left( {p_D^{\left( n \right)}{\gamma _{BB}} + \sigma _U^2} \right) \hfill,  
\end{aligned} 
\end{equation}
\begin{equation}
\begin{aligned}
{f_4}\left( {{p_U}} \right) \leqslant {f_4}\left( {\left. {{p_U}} \right|p_U^{\left( n \right)}} \right) &= \frac{1}{{\text{ln2}}} \cdot \frac{{{\gamma _{3}}}}{{p_U^{\left( n \right)}{\gamma _{3}} + \sigma _D^2}}\left( {{p_U} - p_U^{\left( n \right)}} \right) \hfill \\
&\quad+ {\log _2}\left( {p_U^{\left( n \right)}{\gamma _{3}} + \sigma _D^2} \right) \hfill.  
\end{aligned} 
\end{equation}

Therefore, the solution ${p_D^{\left( n+1 \right)}}$ and ${p_U^{\left( n+1 \right)}}$ can be obtained by solve the following problem
\begin{subequations}\label{opt4}
	\begin{alignat}{2} 
	\mathcal{P}3:& {\mathop {\max}\limits_{{{p_U},{p_D}}}} \tilde R({p_U},{p_D}{\rm{| }}p_U^{(n)},p_D^{(n)}) - \alpha^{(n)} {P_{tot}}({p_U},{p_D}) \label{opt4A} \\
	 \quad{\textrm {s.t.}}
	&\quad\text{8(b)-8(d)},\label{opt4B}
	\end{alignat}
\end{subequations}
where $\tilde R\left( {\left. {{p_D},{p_U}} \right|p_D^{\left( n \right)},p_U^{\left( n \right)}} \right) = {f_1}\left( {{p_D},{p_U}} \right) + {f_2}\left( {{p_D},{p_U}} \right)  - {f_3}\left( {\left. {{p_D}} \right|p_D^{\left( n \right)}} \right) - {f_4}\left( {\left. {{p_U}} \right|p_U^{\left( n \right)}} \right)$ and thus (13a) is global lower bound of (9a).

Problem $\mathcal{P}3$ is a convex optimization problem, and thus CVX can be used to obtain the optimal solution. 
According to \cite{ref13}, $\alpha$ is non-decreasing after each iteration, which means that the system EE will be improved with iterations. Hence, we can update ${p_U}$, ${p_D}$ and $\alpha$ iteratively until the fractional increase of $\alpha$ is below a threshold $\varepsilon_{1}$. The algorithm to solve problem $\mathcal{P}2$ is shown in Algorithm 1.

\begin{algorithm}[h]
	\caption{Dinkelbach’s Method for Problem $\mathcal{P}2$}
	\begin{algorithmic}[1]
		\State {\textbf{Initialize:}} Set initial point \{$p_U^{(0)}$, $p_D^{(0)}$\}, iteration index $n = 0$ and convergence accuracy $\varepsilon_{1}$.
		\Repeat
		\State Update ${\bm{\alpha} ^{(n)}}$ according to (10).
		\State Solve $\mathcal{P}3$ to get $p_U^{(n+1)}$ and $p_D^{(n+1)}$ with given ${\bm{\alpha} ^{(n)}}$.
		\State Update $n=n+1$.
		\Until The fractional increase of ${\bm{\alpha}}$ is below $\varepsilon_{1}$.
		\State \textbf{Output:} Transmit power ${p_U^{*}}$ and $p_D^{*}$.
	\end{algorithmic}
\end{algorithm}

\subsection{Passive Beamforming Optimization With Given ${p_D}$ and ${p_U}$}
To facilitate formulation, we define  ${{\bf{Q}}_t} = {{\bf{q}}_t}{\bf{q}}_t^H$, ${{\bf{Q}}_r} = {{\bf{q}}_r}{\bf{q}}_r^H$, ${{\bf{H}}_1} = {{\bf{h}}_1}{\bf{h}}_1^H$, ${{\bf{H}}_2} = {{\bf{h}}_2}{\bf{h}}_2^H$, and ${{\bf{H}}_3} = {{\bf{h}}_3}{\bf{h}}_3^H$. Then, we can rewrite achieveable sum rate as 
\begin{equation}
R\left( {{{\rm{Q}}_t},{{\bf{Q}}_r}} \right)={{g_1}\left( {{{\bf{Q}}_t}} \right)}+{{g_2}\left( {{{\bf{Q}}_r}} \right)}-{{g_3}\left( {{{\bf{Q}}_t}} \right)}-f
\end{equation}
where ${{g_1}\left( {{{\bf{Q}}_t}} \right)}={{{\log }_2}\left( {{p_U}{\rm{Tr}}\left( {{{\bf{H}}_3}{{\bf{Q}}_t}} \right) + \sigma _D^2 + {p_D}{\rm{Tr}}\left( {{{\bf{H}}_2}{{\bf{Q}}_t}} \right)} \right)}$, ${{g_2}\left( {{{\bf{Q}}_r}} \right)}={{{\log }_2}\left( {{p_D}{{\left| {{h_{BB}}} \right|}^2} + \sigma _U^2 + {p_U}{\rm{Tr}}\left( {{{\bf{H}}_1}{{\bf{Q}}_r}} \right)} \right)}$, ${{g_3}\left( {{{\bf{Q}}_t}} \right)}={{{\log }_2}\left( {{p_U}{\rm{Tr}}\left( {{{\bf{H}}_3}{{\bf{Q}}_t}} \right) + \sigma _D^2} \right)}$ and $f={\log _2}\left( {{p_D}{{\left| {{h_{BB}}} \right|}^2} + \sigma _U^2} \right)$.
%\begin{equation}
%\begin{gathered}
%R\left( {{{\bf{Q}}_t},{{\bf{Q}}_r}} \right) = \underbrace {{{\log }_2}\left( {{p_U}{\text{Tr}}\left( {{{\bf{H}}_3}{{\bf{Q}}_t}} \right) + \sigma _D^2 + {p_D}{\text{Tr}}\left( {{{\bf{H}}_2}{{\bf{Q}}_t}} \right)} \right)}_{{g_1}\left( {{{\bf{Q}}_t}} \right)} \hfill \\
%+ \underbrace {{{\log }_2}\left( {{p_D}{{\left| {{h_{BB}}} \right|}^2} + \sigma _U^2 + {p_U}{\text{Tr}}\left( {{{\bf{H}}_1}{{\bf{Q}}_r}} \right)} \right)}_{{g_2}\left( {{{\bf{Q}}_r}} \right)} \hfill \\
%- \underbrace {{{\log }_2}\left( {{p_U}{\text{Tr}}\left( {{{\bf{H}}_3}{{\bf{Q}}_t}} \right) + \sigma _D^2} \right)}_{{g_3}\left( {{{\bf{Q}}_t}} \right)} - {\log _2}\left( {{p_D}{{\left| {{h_{BB}}} \right|}^2} + \sigma _U^2} \right) \hfill \\ 
%\end{gathered} 
%\end{equation}

With given $p_{U}$ and $p_{D}$, the passive beamforming design subproblem can be written as follows
\begin{subequations}\label{opt5}
	\begin{alignat}{2} 
	\mathcal{P}4:& {\mathop {\max}\limits_{{{{\bf{Q}}_t},{{\bf{Q}}_r},{{\bm{\upbeta }}_t},{{\bm{\upbeta }}_r}}}} R\left( {{{\bf{Q}}_t},{{\bf{Q}}_r}} \right) \label{opt5A} \\
 {\textrm {s.t.}}
&\quad{\rm{rank}}({{\bf{Q}}_l}) = 1,{\rm{ }}\forall l \in \{ t,r\},\label{opt5B}\\
&\quad{\rm{diag}}({{\bf{Q}}_l}) = {{\bm{\upbeta }}_l},{\rm{ }}\forall l \in \{ t,r\},\label{opt5C}\\	&\quad{{\bf{Q}}_l} \succeq 0, \forall l \in \{ t,r\},\label{opt5D}\\
&\quad{p_U}{\rm{Tr(}}{{\bf{Q}}_r}{{\rm{H}}_1}{\rm{)}} \ge R_U^{{\rm{TH}}}({p_D}{\left| {{{\rm{h}}_{{\rm{BB}}}}} \right|^2} + \sigma _U^2),\label{opt5E}\\
&\quad{p_D}{\rm{Tr(}}{{\bf{Q}}_t}{{\rm{H}}_2}{\rm{)}} \ge R_U^{{\rm{TH}}}({p_U}{\rm{Tr}}({{\bf{Q}}_t}{{\rm{H}}_3}) + \sigma _D^2),\label{opt5F}\\
&\quad0 \le \beta _t^m,\beta _r^m \le 1,\beta _t^m + \beta _r^m = 1, m = 1, \ldots ,M.\label{opt5G}
	\end{alignat}
\end{subequations}

Problem $\mathcal{P}4$ is difficult to solve due to the non-convex rank-one constraint. According to \cite{ref2}, for the positive semidefinite matrix ${\bf{Q}}_l$, we have
\begin{equation}
{\rm{rank}}({{\bf{Q}}_l}) = 1 \Leftrightarrow \text{Tr}\left( {{{\bf{Q}}_l}} \right) - \uplambda \left( {{{\bf{Q}}_l}} \right) = 0 ,\forall l \in \{ t,r\}.
\end{equation}

Thus, we can drop the rank-one constraint and use penalty-based method to solve $\mathcal{P}4$, which can be expressed as 
\begin{subequations}\label{opt6}
	\begin{alignat}{2} 
	\mathcal{P}4^{\prime}:\quad& {\mathop {\max}\limits_{{{{\bf{Q}}_t},{{\bf{Q}}_r},{{\bm{\upbeta }}_t},{{\bm{\upbeta }}_r}}}} &\ &F\left( {{{\bf{Q}}_t},{{\bf{Q}}_r}} \right) - G\left( {{{\bf{Q}}_t},{{\bf{Q}}_r}} \right) \label{opt6A} \\
	& \quad{\textrm {s.t.}}
	&&\text{(15c)-(15g)},\label{opt6B}
	\end{alignat}
\end{subequations}
where $F\left( {{{\bf{Q}}_t},{{\bf{Q}}_r}} \right) = {g_1}\left( {{{\bf{Q}}_t}} \right) + {g_2}\left( {{{\bf{Q}}_r}} \right) -f- \frac{1}{\mu }\sum\limits_l {{\text{Tr}}\left( {{{\bf{Q}}_l}} \right)}$ and $G\left( {{{\bf{Q}}_t},{{\bf{Q}}_r}} \right) = {g_3}\left( {{{\bf{Q}}_t}} \right) - \frac{1}{\mu }\sum\limits_l {\uplambda \left( {{{\bf{Q}}_l}} \right)}$. Note that ${\mu }$ is the penalty factor. When $\mu \to 0$, the solution ${{\bf{Q}}_l}$ of $\mathcal{P}4^{\prime}$ always satisfies equation (16), which means that $\mathcal{P}4^{\prime}$ and $\mathcal{P}4$ are equivalent \cite{ref2}. However, if the initial value of the  ${\mu }$ is chosen too small, the objective function of $\mathcal{P}4^{\prime}$ is dominated by the penalty term, which impacts the optimization of original problem. Hence, we first initialize ${\mu }$ with a large value and then gradually reduce ${\mu }$ to a sufficiently small value to eventually obtain the feasible rank-one matrices. 

It can be observed that the objective function in $\mathcal{P}4^{\prime}$ is not a concave one. However, it belongs to the class of difference of two concave functions (DC). Therefore, we can get the upbound of $G\left( {{{\bf{Q}}_t},{{\bf{Q}}_r}} \right)$ as
\begin{equation}
\begin{gathered}
G\left( {{{\bf{Q}}_t},{{\bf{Q}}_r}} \right) \leqslant G\left( {\left. {{{\bf{Q}}_t},{{\bf{Q}}_r}} \right|{\bf{Q}}_t^{\left( i \right)},{\bf{Q}}_t^{\left( i \right)}} \right) \hfill \\
= {\log _2}\left( {{p_U}{\text{Tr}}\left( {{{\bf{H}}_3}{\bf{Q}}_t^{\left( i \right)}} \right) + \sigma _D^2} \right) + \frac{{{p_U}{\text{Tr}}\left[ {{{\bf{H}}_3}\left( {{{\bf{Q}}_t} - {\bf{Q}}_t^{\left( i \right)}} \right)} \right]}}{{\text{ln2}\left[ {{p_U}{\text{Tr}}\left( {{{\bf{H}}_3}{\bf{Q}}_t^{\left( i \right)}} \right) + \sigma _D^2} \right]}} \hfill \\
\;\;\;- \frac{1}{\mu }\sum\limits_l {\left\{ {\uplambda \left( {{\bf{Q}}_l^{\left( i \right)}} \right) + \text{Tr}\left\{ {\operatorname{Re} \left[ {\partial _{{\bf{Q}}_l^{\left( {i} \right)}}^H\uplambda\left( {{{\bf{Q}}_l}} \right)\left( {{{\bf{Q}}_l} - {\bf{Q}}_l^{\left( {i} \right)}} \right)} \right]} \right\}} \right\}} \hfill, \\ 
\end{gathered} 
\end{equation}
where ${\bf{Q}}_l^{\left( {i} \right)}$ is the solution at the  $i$-th iteration. ${\partial _{{\bf{Q}}_l^{\left( {i} \right)}}}\uplambda\left( {{{\bf{Q}}_l}} \right) = {\bm{\sigma}} \left( {{\bf{Q}}_l^{\left( {i} \right)}} \right){\bm{\sigma}} {\left( {{\bf{Q}}_l^{\left( {i} \right)}} \right)^H}$ is the subgradient of $\uplambda ({\bf{Q}}_{l})$ at ${\bf{Q}}_l^{\left( {i} \right)}$ and ${\bm{\sigma}} \left( {{\bf{Q}}_l^{\left( {i} \right)}} \right)$ is eigenvector of the largest eigenvalue \cite{ref2}.

By adopting SCA method \cite{ref14}, the solution at the $(i+1)$-th iteration ${{\bf{Q}}_l^{(i+1)}}$ can be obtained by solving the following problem
\begin{subequations}\label{opt7}
	\begin{alignat}{2} 
	\mathcal{P}5:\quad& {\mathop {\max}\limits_{{{{\bf{Q}}_t},{{\bf{Q}}_r},{{\bm{\upbeta }}_t},{{\bm{\upbeta }}_r}}}} &\ &F\left( {{{\bf{Q}}_t},{{\bf{Q}}_r}} \right) - G\left( {\left. {{{\bf{Q}}_t},{{\bf{Q}}_r}} \right|{\bf{Q}}_t^{\left( i \right)},{\bf{Q}}_r^{\left( i \right)}} \right) \label{opt7A} \\
	& \quad{\textrm {s.t.}}
	&&\text{(15c)-(15g)},\label{opt7B}
	\end{alignat}
\end{subequations}

With given ${\mu }$, $\mathcal{P}5$ is a standard SDP that can be solved by CVX. A two loops algorithm is used to optimize passive beamforming, which is summarized in Algorithm 2. In the inner loop, with given ${\mu }$, we use SCA to solve $\mathcal{P}6$ iteratively until convergence. In the outer loop, the penalty factor is gradually dcreased from one iteration to the next as follows: ${\mu=c\mu }$, where $c<1$. The algorithm terminates when the penalty term satisfies ${\rm{Tr}}({{\bf{Q}}_l}) - \uplambda ({{\bf{Q}}_l})  \le  \varepsilon_{3} ,\forall l \in \{ t,r\}$. Then, we can get the passive beamforming vectors ${{\bf{q}}_t^{*}}$ and ${{\bf{q}}_r^{*}}$ by eigenvalue decomposition.

\begin{algorithm}[h]
	\caption{Penalty-based Algorithm for Problem $\mathcal{P}4$}
	\begin{algorithmic}[1]
		\State {\textbf{Initialize:}} Set initial point \{${{\bf{Q}}_t^{(0)}}$, ${{\bf{Q}}_r^{(0)}}$\}, the penalty factor $\mu$, convergence accuracy $\varepsilon_{2}$ and $\varepsilon_{3}$.
		\Repeat: {\bf{outer loop}}
		\State Set iteration index $i=0$ for inner loop.
			\Repeat: {\bf{inner loop}}
				\State Solve problem $\mathcal{P}5$ to obtain ${{\bf{Q}}_t^{(i+1)}}$ and ${{\bf{Q}}_r^{(i+1)}}$.
				\State Update $i=i+1$.
			\Until the fractional increase of the objective function value is below a predefined threshold $\varepsilon_{2}$.
			\State Update $\mu=c\mu$.
		\Until The solution meet ${\rm{Tr}}({{\bf{Q}}_l}) - \uplambda ({{\bf{Q}}_l}) \le \varepsilon_{3} ,\forall l \in \{ t,r\}$.
		\State \textbf{Output:} ${{\bf{Q}}_t^{*}}$ and ${{\bf{Q}}_r^{*}}$.
	\end{algorithmic}
\end{algorithm}
\subsection{Overall Algorithm and Analysis}
The overall algorithm is implemented by adopting Algorithm 1 and Algorithm 2 iteratively under the AO framework. Similar to \cite{ref13}, the Algorithm 1 will generate a non-decreasing $\alpha$ sequence until convergence with iterations, which means that the system EE will be improved by optimizing the transmit power. When the rank-one constraint is satisfied as the penalty factor decreases, the Algorithm 2 based on SCA will converges to a stationary point of the original problem \cite{ref14}. Then, our AO algorithm will generate a monotonically increasing (at least non-decreasing) sequence. Hence, our proposed algorithm is guaranteed to converge because EE have a upper bound due to the limited resources in the system.

Let $L$, $L_{P}$, $L_{O}$ and $L_{I}$ be the numbers of iterations for AO algorithm, Algorithm 1, outer loop and inner loop of Algorithm 2, respectively. According to \cite{ref15}, the computational complexity of the proposed algorithm is on the order of ${\cal{O}}(L(2{L_P} + 2{L_{{O}}}{L_{{I}}}{M^{3.5}}))$.
\section{SIMULATION RESULTS}
In this section, we evaluate the performance of the STAR-RIS aided FD system with our proposed algorithm. We assume that the locations of BS, STAR-RIS, UL user and DL user are (5m, 45m), (0m,50m), (0m, 35m) and (0m, 100m) respectively. We set $\sigma _U^2 = \sigma _D^2 =  - 90$dBm. The large-scale fading is modelled by $PL\left( d \right) = P{L_0}{\left( {d/{d_0}} \right)^{ - \varpi }}$,  where $P{L_0} =  - 30$dB is the path loss at the reference distance ${d_0} = 1$m, $d$ is the distance, and $\varpi $ is the path-loss exponent which is set to 2.2 \cite{ref2}. We set the Rician factor of communications channels to be 3dB \cite{ref2}. Other required parameters are set as follows unless specified otherwise: ${R_U^{\text{th}}}=1\text{bps/Hz}$, ${R_D^{\text{th}}}=3\text{bps/Hz}$, $\xi=0.1$ \cite{ref11}, $P_{c}=30\text{dBm}$ \cite{ref16}, $P_{c0}=50\text{mW}$ \cite{ref11}, $P_{s}=6\text{dBm}$, $p_D^\text{max}=30\text{dBm}$, $p_U^\text{max}=20\text{dBm}$, $\rho=0.8$, $\sigma _{{\rm{SI}}}^2=-100\text{dB}$\cite{ref10}, $\mu=100$, $c=0.7$, $\varepsilon_{1}=10^{-5}$, $\varepsilon_{2}=10^{-5}$ and $\varepsilon_{3}=10^{-7}$.

The proposed the STAR-RIS aided FD system with EE maximization (SR-FD-EEM), is compared to the following benchmark schemes:
\begin{itemize}
	\item STAR-RIS aided HD system with EE maximization (SR-HD-EEM): We assume that the UL and DL communication is assigned with equal time slot.
	\item Conventional RIS scheme aided FD system with EE maximization (CR-FD-EEM): We assume STAR-RIS operate in conventional RIS scheme, where $M/2$ elements operate in T mode and $M/2$ elements operate in R mode \cite{ref2}.
	\item STAR-RIS aided FD system with sum rate maximization (SR-FD-SRM): We maximize the sum rate of system regardless of the power consumption, which is a special case of our proposed algorithm when $\alpha$ is set as 0 for Algorithm 1.
\end{itemize}
\begin{figure}%[!t]
	\centering
	\includegraphics[width=2.7in]{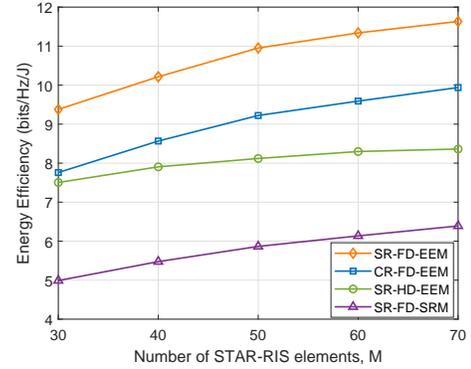}
	% where an .eps filename suffix will be assumed under latex,
	% and a .pdf suffix will be assumed for pdflatex; or what has been declared
	% via \DeclareGraphicsExtensions.
	\caption{EE versus the number of STAR-RIS elements.}
\end{figure}
\begin{figure}%[!t]
	\centering
	\includegraphics[width=2.7in]{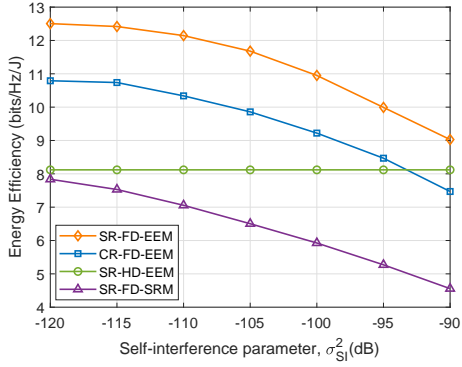}
	% where an .eps filename suffix will be assumed under latex,
	% and a .pdf suffix will be assumed for pdflatex; or what has been declared
	% via \DeclareGraphicsExtensions.
	\caption{EE versus power of the residual SI.}
\end{figure}

Fig. 2 investigates the EE versus different number of elements. It can be seen from the figure that the EE of all sechems increases with $M$. The reason is that larger $M$ significantly improve spectrum efficiency but only slightly increase power consumption, resulting in higher EE. SR-FD-EEM secheme achieves the best EE performance due to the advantages of combining the FD and STAR-RIS technologies. The performance loss of CR-FD-EEM is due to their inflexibility in choosing between transmission and reflection.

The impact of RSI power on the EE performance is presented in Fig. 3. It is observed that EE of FD schemes degrades due to the increasing RSI power while that of HD scheme remains the same. Hence, the powerful SIC is important for FD system.

Fig. 4 shows the EE versus different maximum uplink transmit power. The EE of SR-FD-SRM increases first and then decreases with the increase of $p_U^{\max }$. This is because this scheme maximizes spectral efficiency regardless of power consumption, which cause the loss of EE performance. However, the EE of all schemes with EE maximization first grows monotonically by increasing $p_U^{\max }$ and then reaches to a maximum value due to proper algorithm design. Note that SR-FD-EEM achieves the highest EE compared to other schemes.
\begin{figure}%[!t]
	\centering
	\includegraphics[width=2.7in]{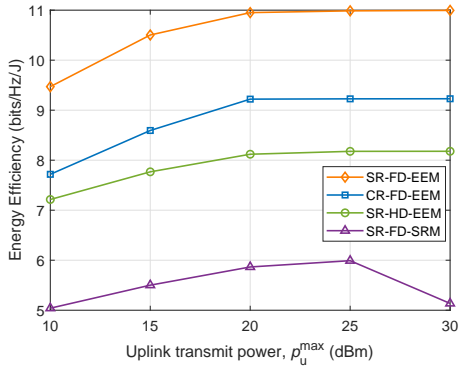}
	% where an .eps filename suffix will be assumed under latex,
	% and a .pdf suffix will be assumed for pdflatex; or what has been declared
	% via \DeclareGraphicsExtensions.
	\caption{EE versus maximum uplink transmit power.}
\end{figure}

Fig. 5 compares the EE of different schemes versus the circuit power of each element at the STAR-RIS. It shows that the EE decreases with the increment of $P_{s}$. This is because the energy consumption of system is increased with the increment of $P_{s}$. Note that SR-FD-EEM outperforms other benchmark schemes, which reveal the benefits of combining FD and STAR-RIS technologies.
\begin{figure}%[!t]
	\centering
	\includegraphics[width=2.7in]{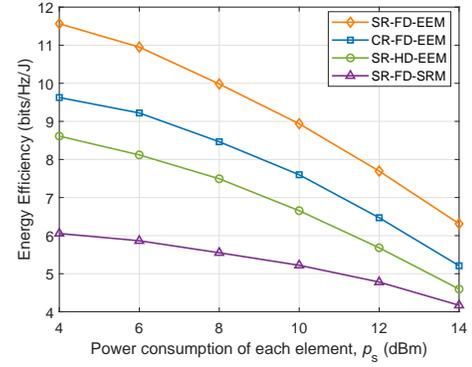}
	% where an .eps filename suffix will be assumed under latex,
	% and a .pdf suffix will be assumed for pdflatex; or what has been declared
	% via \DeclareGraphicsExtensions.
	\caption{EE versus power consumption of each element.}
\end{figure}
\section{CONCLUSION}
This letter studied the EE for a novel STAR-RIS aided FD system. To tackle the non-convex problem, we decouple it into transmit power optimization and passive beamforming design subproblems, and solve them iteratively based on AO. Simulation results show the EE of the STAR-RIS aided FD system is superior to other bechmarks, which reveal the benefits of combining FD and STAR-RIS technologies.

\ifCLASSOPTIONcaptionsoff
  \newpage
\fi

\bibliographystyle{IEEEtran}
\bibliography{IEEEabrv,myre}

\end{document}